\documentclass[prd,twocolumn,showpacs,preprintnumbers,amsmath,amssymb,superscriptaddress,floatfix,nofootinbib]{revtex4}

\usepackage{graphicx}
\usepackage{amsmath}
\usepackage{amsfonts}
\usepackage{amssymb}
\usepackage{color}

\usepackage[colorlinks, citecolor=blue,anchorcolor=red,menucolor=red, linkcolor=red,filecolor=red,runcolor=red,urlcolor=blue,frenchlinks=red]{hyperref}

\begin{document}%\bibliographystyle{unsrt}
\arraycolsep1.5pt

\title{$\Lambda(1405)$ production in the process $\chi_{c0}(1P)\to \bar{\Lambda}\Sigma\pi$}

\author{Li-Juan Liu}
\affiliation{School of Physics and Engineering, Zhengzhou University, Zhengzhou, Henan 450001, China}

\author{En Wang}\email{wangen@zzu.edu.cn}
\affiliation{School of Physics and Engineering, Zhengzhou University, Zhengzhou, Henan 450001, China}

\author{Ju-Jun Xie} \email{xiejujun@impcas.ac.cn}
\affiliation{Institute of Modern Physics, Chinese Academy of
Sciences, Lanzhou 730000, China}

\author{Kai-Lan Song}
\affiliation{School of Physics and Engineering, Zhengzhou University, Zhengzhou, Henan 450001, China}

\author{Jing-Yu Zhu}
\affiliation{School of Physics and Engineering, Zhengzhou University, Zhengzhou, Henan 450001, China}

\date{\today}

\begin{abstract}
We have performed a theoretical study on the process
$\chi_{c0}(1P)\to \bar{\Lambda}\Sigma\pi$, by taking into account
the final state interactions of $\pi\Sigma$ and $\pi\bar{\Lambda}$
based on the chiral unitary approach. As the isospin filters of
$I=0$ in the $\pi\Sigma$ channel and $I=1$ in the $\pi\bar{\Lambda}$
channel, this process can be used to study the molecular structure
of the $\Lambda(1405)$ resonance, and to test the existence of the
predicted states $\Sigma(1380)$ and $\Sigma(1430)$ with spin-parity
$J^P=1/2^-$. Our results show that there is a peak around $1350 \sim
1400$~MeV, and a cusp around the $\bar{K}N$ threshold in the
$\pi\Sigma$ invariant mass distribution, which should be the
important feature of the molecular state $\Lambda(1405)$. We also
find a peak around $1380$~MeV, and a cusp around $\bar{K}N$
threshold in the $\pi\bar{\Lambda}$ invariant mass distribution,
which are associated to the $\Sigma(1380)$ and $\Sigma(1430)$
resonances.
\end{abstract}

\maketitle

\section{Introduction}
\label{sec:introduction}

The nature of the baryon resonances is one of the important issues in hadron physics~\cite{Klempt:2009pi,Crede:2013sze}. There are  many facilities, such as BESIII, LHCb, Belle, {\it et al.}, that have presented a lot of information about baryon resonances, which provides a good platform to extract the baryon properties. On the other hand, the theoretical works go parallel,  most of the existing states can be well described, and some predictions of the effective theories have been confirmed experimentally.

One of the important predictions is the existence of two poles of the $\Lambda(1405)$ state, which were first reported
in Ref.~\cite{Oller:2000fj}, discussed in detail in Ref.~\cite{Jido:2003cb} and later on confirmed in all theories using the chiral unitary approach~\cite{Kaiser:1995eg,Oset:1997it,Hyodo:2002pk,GarciaRecio:2002td,GarciaRecio:2005hy,Borasoy:2005ie,Oller:2006jw,
Borasoy:2006sr,hyodonew,kanchan,hyodorev,ollerguo,maimeissner}.
The experimental confirmation came from the work of Ref.~\cite{Prakhov:2004an} and the analysis of Ref.~\cite{Magas:2005vu}, but other experiments have come to
confirm it too (see Refs.\cite{Roca:2013av,PDG2016} for more details).
The low pole mainly couples to the $\pi\Sigma$ channel, and the high one couples strongly to the $\bar{K}N$ channel~\cite{Roca:2013av,Roca:2013cca}. In addition, many theoretical works about $\Lambda(1405)$ production induced by photon~\cite{Roca:2013av,Roca:2013cca,Wang:2016dtb,Bayar:2017svj,Nacher:1998mi,Nakamura:2013boa}, pion~\cite{Hyodo:2003jw}, kaon~\cite{Nacher:1999ni,Magas:2005vu,Jido:2009jf,Miyagawa:2012xz,Jido:2012cy,Ohnishi:2015iaq}, neutrino~\cite{Ren:2015bsa}, proton-proton collision~\cite{Geng:2007vm,Siebenson:2013rpa,Bayar:2017svj}, and heavy meson decay~\cite{Miyahara:2015cja,Roca:2015tea,Ikeno:2015xea,Xie:2017gwc,Dai:2018hqb,Xie:2018gbi} were carried out to clarify the molecular nature of the  $\Lambda(1405)$ resonance. However, it is problematic to extract the $\Lambda(1405)$ in some conventional reactions involving $\bar{K}N$ or $\pi\Sigma$ final states which may mix $I=0$ and $I=1$ contributions.

In order to further understand the molecular structure of the $\Lambda(1405)$ state,
we investigate the process $\chi_{c0}(1P)\to \bar{\Lambda}\Sigma\pi$, by considering the final state interaction of $\pi\Sigma$, which will dynamically generate the two poles of the $\Lambda(1405)$ state in the chiral unitary approach. The $\chi_{c0}(1P)$, with $I^G(J^{PC})=0^+(0^{++})$, is a $c\bar{c}$ state, and blind to SU(3), hence behaving like an SU(3) singlet. Since the outgoing particle $\bar{\Lambda}$ has isospin $I=0$, the $\pi\Sigma$ system must have isospin $I=0$, to combine to the isospin $I=0$ of the $\chi_{c0}(1P)$. Thus, this process is a good filter of isospin that guarantees that the $\pi\Sigma$ system will be in $I=0$. Besides, the lower $\Lambda(1405)$ pole couples strongly to the $\pi\Sigma$ channel, so the $\pi\Sigma$ final state in this process is an ideal channel to study the molecular structure of the $\Lambda(1405)$ state.

The $\pi$ and $\bar{\Lambda}$ can also undergo final state
interaction, and the isospin of $\pi\bar{\Lambda}$ system is $I=1$,
which together with the $\Sigma$ gives rise to the isospin $I=0$ of
$\chi_{c0}$. It should be noted that a baryon resonance around the
$\bar{K}N$ threshold with $J^P=1/2^-$, strangeness $S=-1$ and
isospin $I=1$ was predicted in the chiral unitary
approach~\cite{Oller:2000fj,Jido:2003cb} (we label this state as
$\Sigma(1430)$ in following), and can couple to the $\pi\Lambda$
channel. It was also suggested to search for this state in the
process $\chi_{c0}\to \bar{\Sigma}\Sigma\pi$ in Ref.~\cite{Wang:2015qta}. In
addition, a $\Sigma^*$ state with $J^P=1/2^-$, mass $M \sim
1380$~MeV and width $\Gamma \sim 120$~MeV (we label this state as
$\Sigma(1380)$ in following), has been predicted in the pentaquark
picture~\cite{Zhang:2004xt}. 
The role of the $\Sigma(1380)$ was investigated
 in the processes of
$J/\psi$ decay~\cite{Zou:2006uh,Zou:2007mk}, $K^-p\to \Lambda
\pi^+\pi^-$~\cite{Wu:2009tu,Wu:2009nw}, $\Lambda p\to \Lambda p
\pi^0$~\cite{Xie:2014zga}, and $\Lambda_c^+\to \eta \pi^+
\Lambda$~\cite{Xie:2017xwx}, as so on. The $\chi_{c0}$ can
also decay into a $\Sigma$ and the intermediate resonance
$\bar\Sigma(1380)$ in $s$-wave, then the $\bar\Sigma(1380)$ decays into
the $\pi\bar{\Lambda}$ states in $s$-wave. As a result, for the
$s$-wave final state interaction of $\pi\bar{\Lambda}$, we will
consider the mechanism of the $\bar\Sigma(1430)$ dynamically generated in the chiral unitary
approach, and the contribution of the intermediate resonance $\bar\Sigma(1380)$
\footnote{We do not consider the contribution of the
intermediate $\bar\Sigma(1385)$ with $J^P=3/2^-$ in
$\pi\bar{\Lambda}$ system, because the
$\bar\Sigma(1385)$, as the SU(3) anti-decuplet, together with the
SU(3) octet state $\Sigma$, cannot give rise to the SU(3) singlet state
$\chi_{c0}$, and its contribution will be suppressed.}. In this way, the shape of the $\pi\bar{\Lambda}$ mass
distribution of this process can be helpful to test the existences of
the $\Sigma(1430)$ and $\Sigma(1380)$ resonances~\cite{Zou:2007mk,PDG2016}. It should be noted that the
$\Sigma(1430)$ and $\Sigma(1380)$ resonances are assumed to be the members of the SU(3)
octet in this work, since the anti-singlet and anti-decuplet
of $\pi\bar\Lambda$ system are suppressed in the $\chi_{c0}\to 
\bar\Lambda\Sigma \pi$ reaction. 

In addition, the Lattice QCD groups have
also calculated the mass of the $\Sigma^*$ with $J^P=1/2^-$. For instance, the mass of the lowest $\Sigma^*$ ($J^P=1/2^-$) is predicted to be about 1.2~GeV in Ref.~\cite{Edwards:2012fx} and $1.6\sim 1.8$~GeV in Refs.~\cite{Engel:2013lea,Engel:2013ig}, both of which are inconsistent with the masses of $\Sigma(1430)$ and $\Sigma(1385)$. Thus, searching for the
$\Sigma^*$ state with $J^P=1/2^-$ experimentally could distinguish the different theoretical predictions.

This paper is organized as follows. In Sec.~\ref{sec:formalism}, we will give the formalism of mechanisms, in Sec.~\ref{sec:result}, we will present our results and discussions, finally, the conclusion will be given in Sec.~\ref{sec:conclusion}.

\section{Formalism}
\label{sec:formalism}

In this section, we will describe the reaction mechanism for the process $\chi_{c0}(1P)\to \bar{\Lambda}\Sigma\pi$.
\subsection{The model of $\chi_{c0}(1P)\to \bar{\Lambda}\Sigma\pi$}

\begin{figure}
\includegraphics[width=0.4\textwidth]{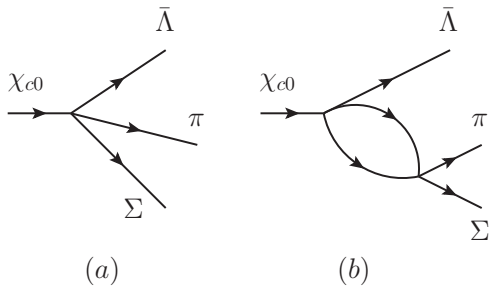}
\caption{Diagrams for $\chi_{c0}\to \bar{\Lambda}\Sigma\pi$ decay: (a) direct $\bar{\Lambda}\Sigma\pi$ vertex at tree level, (b) final state interaction of $\pi\Sigma$.}
\label{fig:feynman}
\end{figure}

In the first step, $\chi_{c0}(1P)$ can decay into the final states $\bar{\Lambda}\Sigma\pi$ directly in the tree level, as depicted in Fig.~\ref{fig:feynman}(a). Then the final particles $\pi\Sigma$ can undergo the final state interaction, which will dynamically generate the resonance $\Lambda(1405)$, as depicted in Fig.~\ref{fig:feynman}(b).

\begin{table}
\caption{SU(3) isoscalar coefficients for the $\left\langle\bar{\Lambda}|MB\right\rangle$ matrix elements.}
\label{tab:weight}
\begin{tabular}{c|cccc}
\hline \hline
$\bar{\Lambda}$ & $\bar{K}N$ & $\pi\Sigma$ & $\eta\Lambda$  & $K\Xi$ \\ \hline
$\tilde{D}$ & $ \sqrt{\frac{1}{10}}$  & $-\sqrt{\frac{3}{5}}$ & $-\sqrt{\frac{1}{5}}$ & $-\sqrt{\frac{1}{10}}$ \\
$\tilde{F}$ & $\sqrt{\frac{1}{2}}$  & 0 & 0 & $\sqrt{\frac{1}{2}}$ \\
\hline
\end{tabular}
\end{table}

For the mechanism of the $\pi\Sigma$ final state interaction, we must take into account that, in the first step, one can produce the other meson-baryon pairs that couple to the same $\pi\Sigma$ quantum numbers, then reaching the final $\pi\Sigma$ through re-scattering. Since $\chi_{c0}$ behaves like an SU(3) singlet, and $\bar{\Lambda}$ is an SU(3) anti-octet, hence, $\pi\Sigma$ system should be in an SU(3) octet, which will give rise to the same quantum numbers as $\Lambda(1405)$. Since both $\pi$ and $\Sigma$ belong to the members of SU(3) octet, we have the representation for $8(\pi)\bigotimes 8(\Sigma)$ going to the $8^s$ (symmetry) and $8^a$ (antisymmetry). With the SU(3) isoscalar factors given in the PDG~\cite{PDG2016}, tabulated in Table~\ref{tab:weight}, we can obtain the weights $h_i$ in the isospin basis of this states, which go into the primary production of each meson-baryon channel, and stand for the weights of the transition $\chi_{c0}\to \bar{\Lambda}\, PB$ ($PB=\bar{K}N,\pi\Sigma,\eta\Lambda,\Xi K$),
\begin{eqnarray}
&&h_{\bar{K}N}=\sqrt{\frac{1}{10}}\tilde{D}+\sqrt{\frac{1}{2}}\tilde{F},~~ h_{\pi\Sigma}=-\sqrt{\frac{3}{5}}\tilde{D},\nonumber \\
&&h_{\eta\Lambda}=-\sqrt{\frac{1}{5}}\tilde{D},~~ h_{K\Xi}=-\sqrt{\frac{1}{10}}\tilde{D}+\sqrt{\frac{1}{2}}\tilde{F},\label{eq:weight_piSigma}
\end{eqnarray}
where $\tilde{D}$ and $\tilde{F}$ are unknown parameters.

The interaction of the octet pseudoscalar mesons and the octet $1/2^+$ baryons, which can dynamically generate the $\Lambda(1405)$, has been studied within the chiral unitary approach in Refs.~\cite{Oset:1997it,Oller:2000fj,Oset:2001cn}. In Ref.~\cite{Roca:2013cca}, a new strategy to extract the positions of the two poles of $\Lambda(1405)$ from $\pi\Sigma$ photoproduction experimental data was done, based on small modifications of the unitary chiral perturbation theory amplitude, which will be adopted in this work. Because the thresholds of the channels $\eta\Lambda$ and $K\Xi$ lay far above the energies that we consider in this work, and their effect can be effectively reabsorbed in the subtraction constants, as discussed in Refs.~\cite{Roca:2013av,Roca:2013cca}, we donot consider these two channels in following calculations.

\begin{figure}
\includegraphics[width=0.3\textwidth]{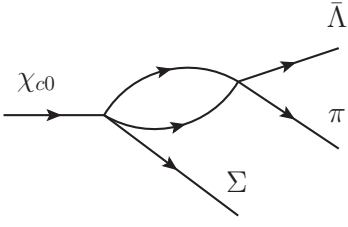}
\caption{The $s$-wave final state interaction of  $\pi\bar{\Lambda}$ in the process $\chi_{c0}\to \bar{\Lambda} \Sigma \pi$.}
\label{fig:piLambda}
\end{figure}

In addition to the $\pi\Sigma$ final state interaction, the states $\pi$ and $\bar{\Lambda}$ also can undergo the final state interaction in the process of $\chi_{c0}\to \bar{\Lambda}\Sigma\pi$, as depicted in Fig.~\ref{fig:piLambda}.
In this case, we get the weights for different channels from Ref.~\cite{Wang:2015qta},
\begin{eqnarray}
&&\tilde{h}_{K\bar{N}}=-\sqrt{\frac{3}{8}}\tilde{D}+\sqrt{\frac{1}{6}}\tilde{F},\nonumber \\
&& \tilde{h}_{\pi\bar{\Sigma}}=\sqrt{\frac{2}{3}}\tilde{F},~~\tilde{h}_{\pi\bar{\Lambda}}=\sqrt{\frac{1}{5}}\tilde{D},~~\tilde{h}_{\eta\bar{\Sigma}}=\sqrt{\frac{1}{5}}\tilde{D},\nonumber \\
&&\tilde{h}_{\bar{K}\bar{\Xi}}=-\sqrt{\frac{3}{8}}\tilde{D}+\sqrt{\frac{1}{6}}\tilde{F},  \label{eq:weight2}
\end{eqnarray}
where the parameters $\tilde{D}$ and $\tilde{F}$ are same as those of Eq.~(\ref{eq:weight_piSigma}).

As predicted in Refs.~\cite{Oller:2000fj,Jido:2003cb}, the $\pi\bar{\Lambda}$ system with isospin $I=1$  can undergo the $s$-wave final state interaction, which will dynamically generate a cusp structure around the $\bar{K}N$ threshold, associated to the $\Sigma(1430)$ resonance.  Taking into account the $s$-wave final state interaction, the total amplitude for the process $\chi_{c0}\to\bar{\Lambda}\Sigma\pi$ is,
\begin{eqnarray}
\mathcal{M} &=& V_p \left[ h_{\pi\Sigma} + \sum_i h_i G_i(M_{\pi\Sigma}) t_{i,\pi\Sigma}(M_{\pi\Sigma}) \right]\nonumber \\
&& +  \tilde{V}_p \sum_j \tilde{h}_j G_j(M_{\pi\bar{\Lambda}}) t_{j,\pi\bar{\Lambda}}(M_{\pi\bar{\Lambda}}) \nonumber \\
&=& \mathcal{M}_{\rm tree}+ \mathcal{M}_{\pi\Sigma}+\mathcal{M}_{\pi\bar{\Lambda}}
\label{eq:amp_full1}
\end{eqnarray}
where the loop function $G$ and the transition amplitude $t$ of the $\pi\Sigma$ final state interaction are taken from Refs.~\cite{Roca:2013cca,Wang:2015qta}. 
$M_{\pi\Sigma}$ and $M_{\pi\bar{\Lambda}}$ are the invariant masses of the $\pi\Sigma$ and $\pi\bar{\Lambda}$, respectively.
In this work, we work with $R=\tilde{F}/\tilde{D}$, and include the weight $\tilde{D}$ of Eqs.~(\ref{eq:weight_piSigma}) and (\ref{eq:weight2}) in the $V_p$ and $\tilde{V}_p$ factors, as done in Ref.~\cite{Wang:2015qta}.

In Eq.(\ref{eq:amp_full1}), the amplitude for the tree level of Fig.~\ref{fig:feynman}(a) is,
\begin{equation}
\mathcal{M}_{\rm tree}=V_p h_{\pi\Sigma}. \label{eq:treeamp1}
\end{equation}
If we consider the transition $\chi_{c0}\to \Sigma P\bar{B}$ ($P\bar{B}=K\bar{N},~\pi\bar{\Sigma},~\pi\bar{\Lambda},~\eta\bar{\Sigma},~\bar{K}\bar{\Xi}$), the amplitude for the tree level can also be given as,
\begin{equation}
\mathcal{M}_{\rm tree}=\tilde{V}_p \tilde{h}_{\pi\bar{\Lambda}}. \label{eq:treeamp2}
\end{equation}
Combing the Eqs.~(\ref{eq:treeamp1}) and (\ref{eq:treeamp2}), we can obtain that $\tilde{V}_p=V_p\times(h_{\pi\Sigma}/\tilde{h}_{\pi\bar{\Lambda}})=-\sqrt{3}V_p$. Now, the amplitude of Eq.~(\ref{eq:amp_full1}) can be rewritten as,
\begin{eqnarray}
\mathcal{M} &=& V_p \left[ h_{\pi\Sigma} + \sum_i h_i G_i(M_{\pi\Sigma}) t_{i,\pi\Sigma}(M_{\pi\Sigma}) \right. \nonumber \\
&&  \left. -\sqrt{3} \sum_j \tilde{h}_j G_j(M_{\pi\bar{\Lambda}}) t_{j,\pi\bar{\Lambda}}(M_{\pi\bar{\Lambda}}) \right] \nonumber \\
&=& V_p \left(h_{\pi\Sigma} + T_{\pi\Sigma} + T_{\pi\bar{\Lambda}}\right). \label{eq:amp2}
\end{eqnarray}

\begin{figure}
\includegraphics[width=0.4\textwidth]{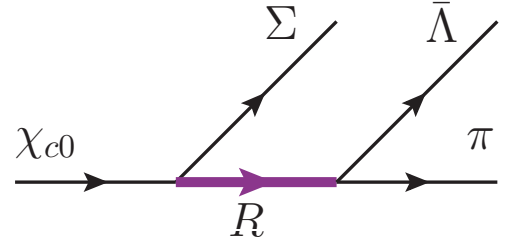}
\caption{(color online) The mechanism of the intermediate resonance $\bar\Sigma(1380)$ in the $\pi\bar{\Lambda}$ channel in the process $\chi_{c0}\to \bar{\Lambda} \Sigma \pi$.}
\label{fig:sigmastar}
\end{figure}

\subsection{Contribution from $\Sigma(1380)$ with $J^P=1/2^-$}
In addition to the $s$-wave interaction, the $\pi\bar{\Lambda}$ states can also couple to $\bar\Sigma(1380)$ state with $J^P=1/2^+$ in $s$-wave, as depicted in Fig.~\ref{fig:sigmastar}.  The contribution of intermediate resonance  $\bar{\Sigma}(1380)$ should add coherently to the Eq.~(\ref{eq:amp2})\footnote{Although the $\Sigma(1380)$ is predicted in pentaquark picture~~\cite{Zhang:2004xt}, for simplicity, we use the Breit-Wigner form for its contribution, as used in Refs.~\cite{Wu:2009tu,Wu:2009nw,Xie:2014zga,Xie:2017xwx}.}, as follows,
\begin{eqnarray}
\mathcal{M} &=&  V_p \left(h_{\pi\Sigma} + T_{\pi\Sigma} + T_{\pi\bar{\Lambda}} + T_{\Sigma(1380)} \right), \label{eq:fullmodel} \\
T_{\Sigma(1380)} &=&  \frac{\alpha M_{\Sigma(1380)}}{M_{\pi\bar{\Lambda}}-M_{{\Sigma}(1380)} + i\Gamma_{{\Sigma}(1380)}/2 },
\label{eq:amppiLambda}
\end{eqnarray}
where the normalization $\alpha$ stands for the amplitude strength, which will be chosen to provide a sizeable effect of the intermediate resonance $\bar{\Sigma}(1380)$. 
In this work, we take $M_{\Sigma(1380)}=1380$~MeV, $\Gamma_{\Sigma(1380)}=120$~MeV as fitted in Ref.~\cite{Wu:2009tu}.

With the amplitude of Eq.~(\ref{eq:fullmodel}), the invariant mass distribution of $\chi_{c0}(1P)\to \bar{\Lambda}\Sigma\pi$ can be written as,
\begin{align}\label{eq:dwidth}
\frac{d\Gamma}{dM^2_{\pi\Sigma}dM^2_{\pi\bar\Lambda}}
=\frac{1}{(2\pi)^3}\frac{4M_\Sigma M_\Lambda}{32M^3_{\chi_{c0}}} \left|\mathcal{M}(M_{\pi\Sigma}, M_{\pi\bar\Lambda})\right|^2.
\end{align}
For a given value of $M^2_{\pi\Sigma}$, the range of $M^2_{\pi\bar\Lambda}$ is defined as,
\begin{eqnarray}
(M^2_{\pi\bar\Lambda})_{\rm max}\!&=&\! \left(E^*_\pi+E^*_{\bar\Lambda}\right)^2-\left(\sqrt{E^{*2}_\pi-m^2_\pi}-\sqrt{E^{*2}_{\bar\Lambda}-m^2_{\Lambda}}\right)^2,  \nonumber \\
(M^2_{\pi\bar\Lambda})_{\rm min}\! &=&\! \left(E^*_\pi+E^*_{\bar\Lambda}\right)^2-\left(\sqrt{E^{*2}_\pi-m^2_\pi}+\sqrt{E^{*2}_{\bar\Lambda}-m^2_{\Lambda}}\right)^2,  \nonumber \\
\end{eqnarray}
here $E^{*}_\pi=(M^2_{\pi\Sigma}-M^2_\Sigma+m^2_\pi)/2M_{\pi\Sigma}$ and $E^{*}_{\bar\Lambda}=(M^2_{\chi_{c0}}-M^2_{\pi\Sigma}-M^2_{\Lambda})/2M_{\pi\Sigma}$ are the energies of $\pi$ and $\bar\Lambda$ in the rest frame of the $\pi\Sigma$ system.
\section{Result and Discussions}
\label{sec:result}

\begin{figure}
\includegraphics[width=0.4\textwidth]{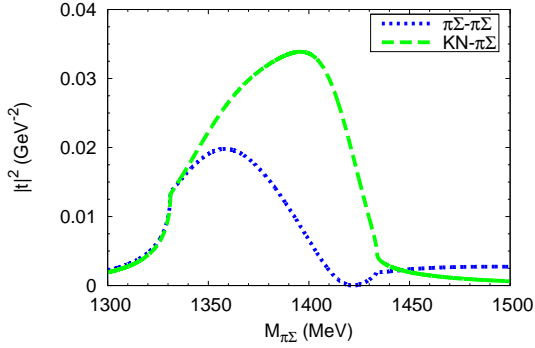}
\caption{(color online) Modulus squared of the amplitudes $|t_{\bar{K}N,\pi\Sigma}|^2$ and  $|t_{\pi\Sigma,\pi\Sigma}|^2$ in $I=0$.}
\label{fig:amp}
\end{figure}

\begin{figure}
\includegraphics[width=0.4\textwidth]{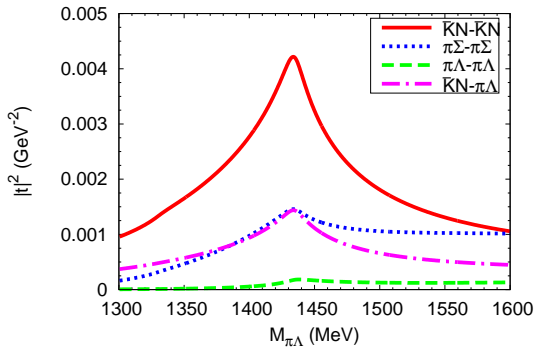}
\caption{(color online) Modulus squared of the amplitudes $|t_{\bar{K}N,\bar{K}N}|^2$, $|t_{\pi\Sigma,\pi\Sigma}|^2$, $|t_{\pi\Lambda,\pi\Lambda}|^2$ and $|t_{\bar{K}N,\pi\Lambda}|^2$.}
\label{fig:amp_piLambda}
\end{figure}

Before presenting the results for the process $\chi_{c0}(1P)\to \bar{\Lambda}\Sigma \pi$, we show the modulus squared of the amplitudes $|t_{\bar{K}N,\pi\Sigma}|^2$ and  $|t_{\pi\Sigma,\pi\Sigma}|^2$ in $I=0$ in Fig.~\ref{fig:amp}, from where we can see that the peak of $\pi\Sigma\to \pi\Sigma$ amplitude mainly comes from the lower pole, while the one of  $\bar{K}N \to \pi\Sigma$ amplitude comes from the higher pole. In Fig.~\ref{fig:amp_piLambda}, we present the modulus squared of the transition amplitudes $|t|^2$ in $I=1$. As we can see, a clear cusp structure around the $\bar{K}N$ threshold is found, same as the Refs.~\cite{Roca:2013cca,Wang:2015qta}.

\begin{figure}
\includegraphics[width=0.4\textwidth]{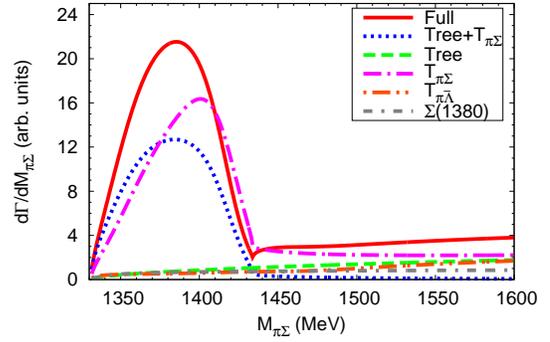}
\caption{(color online)
The $\pi\Sigma$ mass distribution of the process $\chi_{c0}\to \bar{\Lambda}\Sigma \pi$.
 The curve labeled as `Full' shows the  result of all the contributions as given by Eq.~(\ref{eq:fullmodel}), the curves labeled as `Tree', `$T_{\pi\Sigma}$', `$T_{\pi\bar{\Lambda}}$', and `$\bar\Sigma(1380)$' correspond to the contributions of the
tree level [Fig.~\ref{fig:feynman}(a)], the $\pi\Sigma$ final state interaction [Fig.~\ref{fig:feynman}(b)], the contribution of the $\bar\Sigma(1430)$ generated in $s$-wave $\pi\bar{\Lambda}$ final state interaction (Fig.~\ref{fig:piLambda}), and the intermediate resonance $\bar\Sigma(1380)$ (Fig.~\ref{fig:sigmastar}), respectively.
Finally, the curve labeled as `Tree+$T_{\pi\Sigma}$' represents the total contribution of the tree level [Fig.~\ref{fig:feynman}(a)] and the $\pi\Sigma$ final state interaction [Fig.~\ref{fig:feynman}(b)].}
\label{fig:widthpiSigma}
\end{figure}

\begin{figure}
\includegraphics[width=0.4\textwidth]{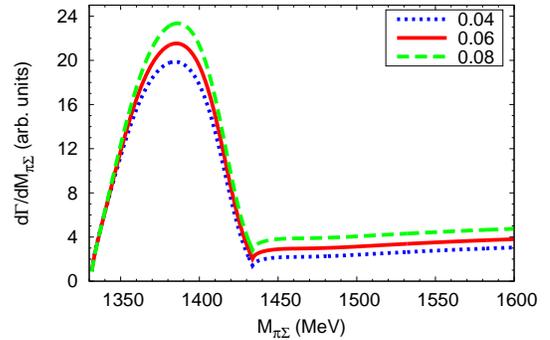}
\caption{(color online)
The $\pi\Sigma$ mass distribution of the process $\chi_{c0}\to \bar{\Lambda}\Sigma \pi$ for different values of $\alpha$ of Eq.~(\ref{eq:amppiLambda}).}
\label{fig:widthpiSigma_alpha}
\end{figure}

As we donot know the exact value of $R(={\tilde{F}}/{\tilde{D}})$, and the production weights of $\bar{K}N$ and $\pi\Sigma$ are expected to be same magnitude, we will vary the $R$ from -2 to 2, as done in Ref.~\cite{Wang:2015qta}.  Firstly, we take $R=1$, and  $\alpha=0.06$ of the normalization in Eq.~(\ref{eq:amppiLambda}), which gives rise to a sizeable effect of intermediate $\bar\Sigma(1380)$. In Fig.~\ref{fig:widthpiSigma}, up to an arbitrary normalization of $V_p$, we show the $\pi\Sigma$ invariant mass distribution, where the term of $\pi\Sigma$ final state interaction (labeled as `$T_{\pi\Sigma}$') gives rise to a peak around $1410$~MeV, and the peak moves to low energy because of the interference with the tree level term (labeled as `tree').
We also present the $\pi\Sigma$  invariant mass distribution with different values of $\alpha$ in Fig.~\ref{fig:widthpiSigma_alpha}, which shows that
the contribution from the intermediate $\bar{\Sigma}(1380)$ resonance does not significantly affect the peak position of the $\pi\Sigma$ mass distribution.

\begin{figure}
\includegraphics[width=0.4\textwidth]{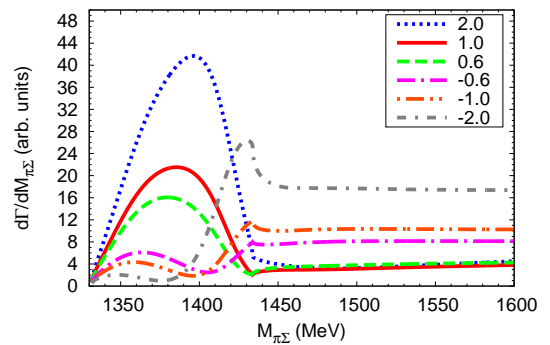}
\caption{(color online) The $\pi\Sigma$ mass distribution of the process $\chi_{c0}\to \bar{\Lambda}\Sigma \pi$ for different values of the ratio $R$.}
\label{fig:widthpiSigma2}
\end{figure}

The $\pi\Sigma$ invariant mass distribution with different values of $R$, varying from -2 to 2, is shown in Fig.~\ref{fig:widthpiSigma2}, where  we can see that as the ratio $R$ decreases, the peak position of the $\Lambda(1405)$ in the $\pi\Sigma$ mass distribution moves to the region of low energies, and a cusp structure around $\bar{K}N$ threshold appears. Indeed, the value of $R$ can be larger than 2 or less than -2, and the peak of the $\pi\Sigma$ invariant mass distribution will vary from 1360~MeV, the position of the peak in $t_{\pi\Sigma,\pi\Sigma}$, to 1400~MeV, the one in $t_{\bar{K}N,\pi\Sigma}$, as shown in Fig.~\ref{fig:amp}.

As discussed in Refs.~\cite{Dai:2018tgo,Wang:2017mrt, Wang:2018djr,Dai:2018nmw}, one of the defining features associated to the molecular states that couple to several hadron-hadron channels is that one can find a strong and unexpected cusp  at the threshold of the channels corresponding to the main component of the
molecular state, and one of the examples is the observations of the cusp, associated to the molecular state $X(4160)$, in the $B^+\to J/\psi \phi K^+$ decay~\cite{Wang:2017mrt}. The peak and the cusp, observed in Fig.~\ref{fig:widthpiSigma2}, should be the important feature to confirm the existence of the $\Lambda(1405)$ in the decay of $\chi_{c0}\to \bar{\Lambda}\Sigma\pi$. Thus, we strongly encourage to measure the $\pi\Sigma$ invariant mass distribution of the $\chi_{c0}\to \bar{\Lambda}\Sigma\pi$ decay.

\begin{figure}
\includegraphics[width=0.4\textwidth]{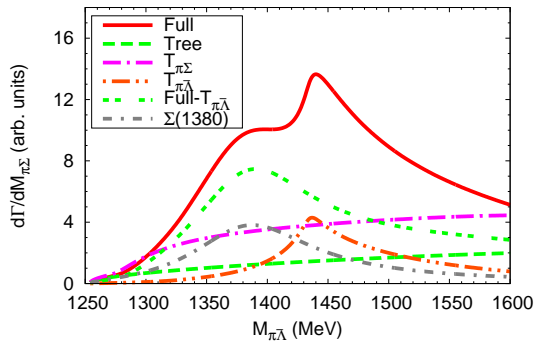}
\caption{(color online) The $\pi\bar{\Lambda}$ mass distribution of the process $\chi_{c0}\to \bar{\Lambda}\Sigma\pi$. The curve labeled as `Full-$T_{\pi\bar{\Lambda}}$' corresponds to our full results excluding the contribution of the $\bar\Sigma(1430)$.  The explanations of the other curves are same as those of Fig.~\ref{fig:widthpiSigma}.}
\label{fig:dwidthpiLambda}
\end{figure}
\begin{figure}
\includegraphics[width=0.4\textwidth]{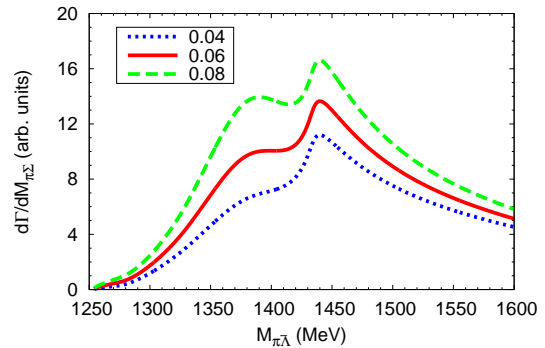}
\caption{(color online)
The $\pi\bar{\Lambda}$ mass distribution of the process $\chi_{c0}\to \bar{\Lambda}\Sigma \pi$ for different values of $\alpha$ of Eq.~(\ref{eq:amppiLambda}).}
\label{fig:widthpiLambda_alpha}
\end{figure}

In order to check the existence of the $\Sigma(1430)$ and $\Sigma(1380)$,
we also present the $\pi\bar{\Lambda}$ invariant mass distribution in Fig.~\ref{fig:dwidthpiLambda}, where we can see that there is a clear bump structure around 1380~MeV, which is associated to the intermediate resonance $\Sigma(1380)$, and a clear cusp structure around the $\bar{K}N$ threshold,  which comes from the $\pi\bar{\Lambda}$ final state interaction, as shown in Eq.~(\ref{eq:amp2}). By varying the value of the normalization $\alpha$ as depicted in Fig.~\ref{fig:widthpiLambda_alpha}, the bump structure of $\bar{\Sigma}(1380)$ becomes smoother for a smaller $\alpha$, and more clear for a larger one. It should be stressed that the bump structure of $\bar{\Sigma}(1380)$, or the cusp structure around the $\bar{K}N$ threshold, if confirmed experimentally, should be related to the resonance $\Sigma(1380)$ or $\Sigma(1430)$.

\begin{figure}
\includegraphics[width=0.4\textwidth]{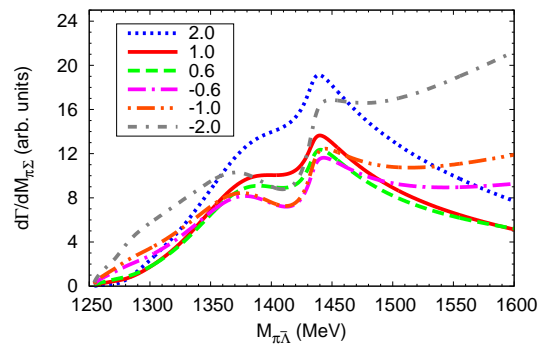}
\caption{(color online) The $\pi\bar{\Lambda}$ mass distribution of the process $\chi_{c0}\to \bar{\Lambda}\Sigma \pi$ for different values of the ratio $R$.}
\label{fig:dwidthpiLambda2}
\end{figure}

In Fig.~\ref{fig:dwidthpiLambda2}, we show the $\pi\bar\Lambda$ invariant mass distributions by varying the ratio $R$ from -2 to 2 . As we can see that the peak structure of the $\bar{\Sigma}(1380)$ and the cusp structure around the $\bar{K}N$ threshold are always clear for different values of the ratio $R$.

\begin{figure}
\includegraphics[width=0.4\textwidth]{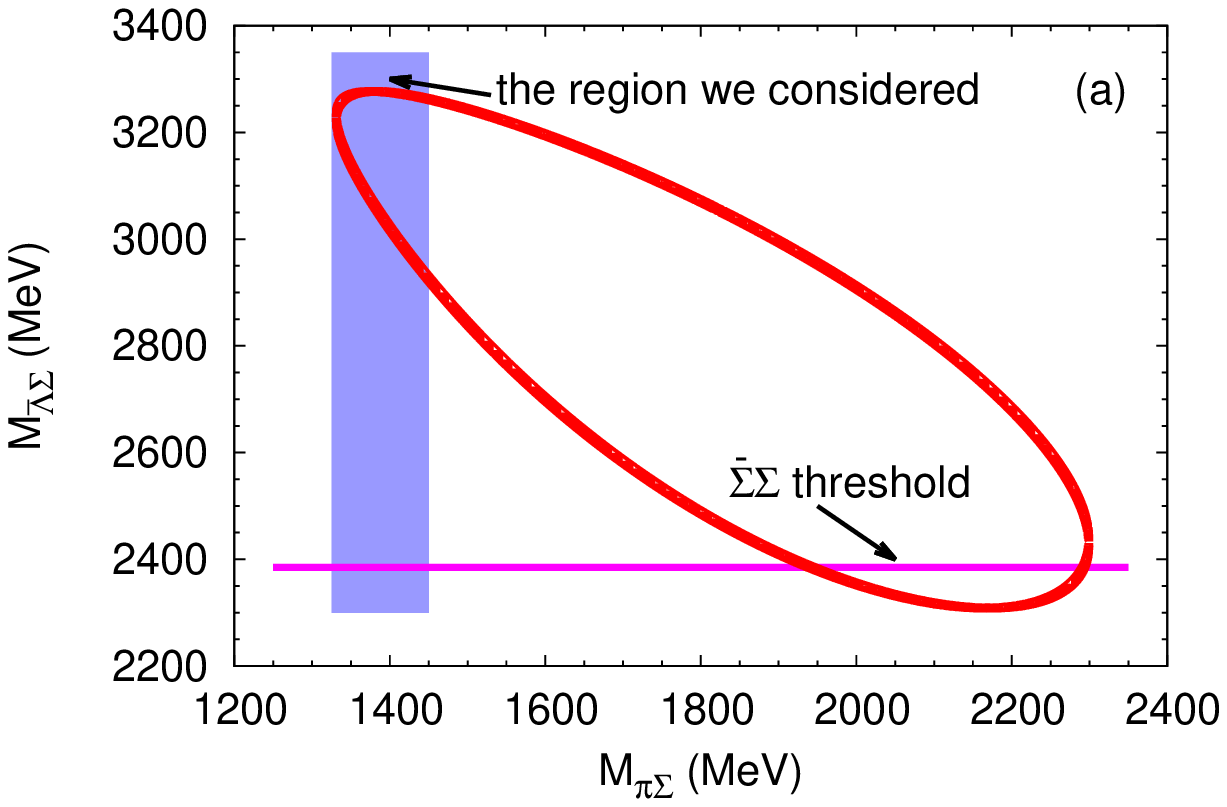}
\includegraphics[width=0.4\textwidth]{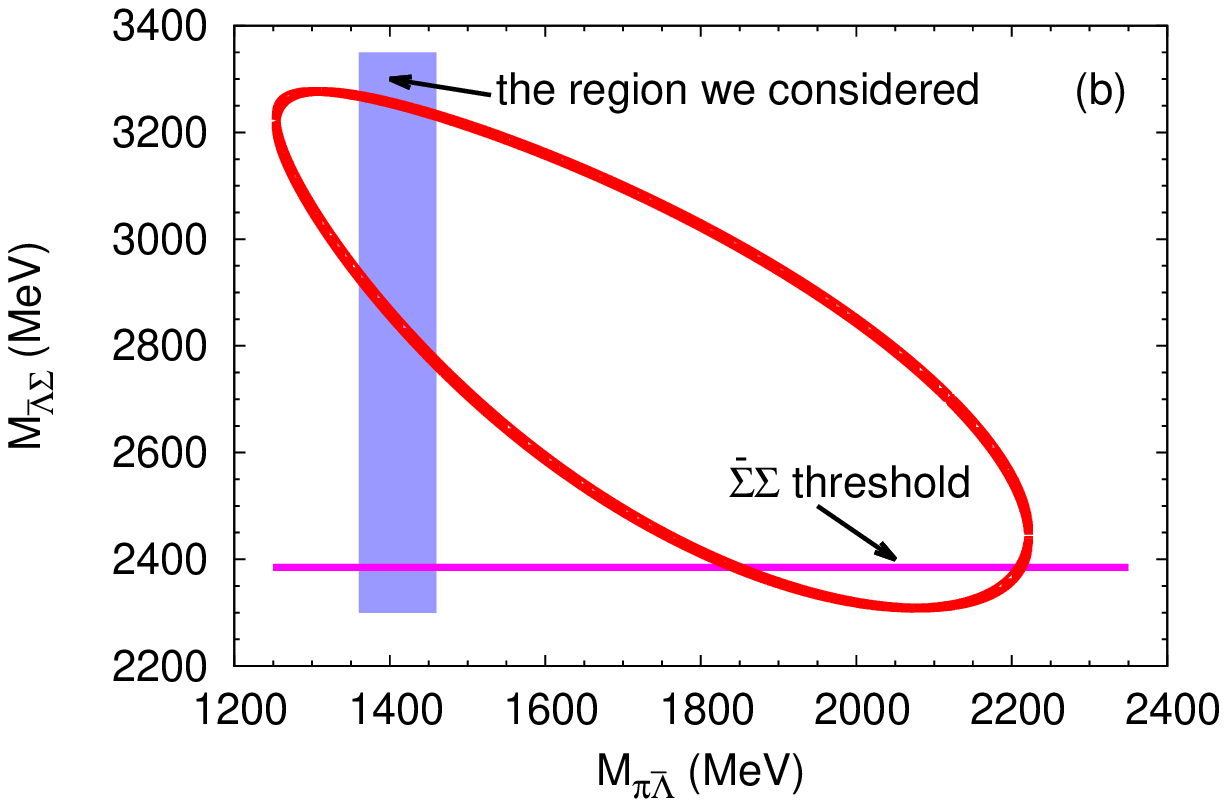}
\caption{(color online) The Dalitz  plots for the $\chi_{c0}\to \bar{\Lambda}\Sigma\pi$ process.}
\label{fig:dalitz}
\end{figure}

In addition, the $\bar{\Lambda}\Sigma$ states can also process the
final state interaction, which could give rise to some sharp
structure (for instances, cusp) at the $\bar{\Sigma}\Sigma$
threshold, and induce some reflection effect on the $\pi\Sigma$ and
$\pi\bar{\Lambda}$ invariant mass distributions.  By looking at the
Dalitz plots of the $\chi_{c0}\to \bar{\Lambda}\Sigma \pi$ process,
as shown in Fig.~\ref{fig:dalitz}, we find that the sharp structure
around $\bar{\Sigma}\Sigma$ in the $\bar{\Lambda}\Sigma$
distribution only contributes to the $\pi\Sigma$ and
$\pi\bar{\Lambda}$ invariant mass distribution $400 \sim 500$~MeV
beyond the regions of the $\Lambda(1405)$ and the $\Sigma^*$.
Indeed, we have discussed the effect of the sharp structure around
$\bar{\Sigma}\Sigma$ threshold on the $\pi\Sigma$ invariant mass
distribution of the $\chi_{c0}\to \bar{\Sigma}\Sigma\pi$ process in
Ref.~\cite{Wang:2015qta}, a similar reaction as the present one, and
show that the structure of $\pi\Sigma$ invariant mass distribution
remains unchanged. Thus, the sharp structure around
$\bar{\Sigma}\Sigma$ threshold should have no significant effect on
our predictions, and can be safely neglected in here.

So far we have considered only two-body interactions of the final
three particles. It is known that the three-body dynamics plays also
a major and sometimes dominant role in the understanding of hadronic
resonances~\cite{Mai:2017vot,Mai:2017bge,Mai:2017wdv,Doring:2018xxx}
However, including such contributions the decay amplitudes would become more complex due to additional parameters, and we cannot
determine or constraint these parameters, since there are no
experimental data on the $\chi_{c0}(1P)\to\bar{\Lambda}\Sigma\pi$
decay. Hence, we will leave these contributions to future studies
when experimental data become available.

Finally, it should be noted that the SIDDHARTA measurement of kaonic hydrogen gives an accurate constraint on the $K^-p$ scattering length~\cite{Bazzi:2012eq}, and the interaction kernel with the next-to-leading order chiral perturbation theory is used with the systematic $\chi^2$ analysis in Refs.~\cite{Mai:2014xna,Mai:2012dt,Mai:2018rjx,Mai:2014uma, hyodonew,hyodorev,ollerguo} (The important theoretical input for a reliable analysis of SIDDHARTA data can be seen in Ref.~\cite{Meissner:2004jr}). However, as a motivation for measuring this process experimentally, we use the model of the final state interaction of the $\pi\Sigma$ of $I=0$ developed in Ref.~\cite{Roca:2013cca}, which can also produce the main feature of the amplitudes of $\pi\Sigma \to \bar{K}N$ and $\pi\Sigma \to \pi\Sigma $ given by the more accurate model.

\section{Conclusions}
\label{sec:conclusion}

In this paper, we have studied the process $\chi_{c0}(1P)\to\bar{\Lambda}\Sigma\pi$ by taking into account the final state interactions of $\pi\Sigma$ and $\pi\bar{\Lambda}$ within the chiral unitary approach. As the isospin $I=0$ filter in the $\pi\Sigma$ system and the isospin $I=1$ filter in the $\pi\bar{\Lambda}$ system, this process can be used to study the molecular structure of $\Lambda(1405)$ state, and to search for the predicted states $\Sigma(1380)$ or $\Sigma(1430)$ with $J^P=1/2^-$.
We have shown that, there is a peak of  $1350 \sim 1400$~MeV, and a cusp around the $\bar{K}N$ in the $\pi\Sigma$ mass distribution, which should be the important feature of the molecular state $\Lambda(1405)$.

Indeed, the intermediate $\bar{\Sigma}(1385)$ state cannot be fully suppressed because SU(3) symmetry is known be substantially broken. If a bump is observed around $1380$ MeV in the $\pi\bar{\Lambda}$ invariant mass distribution, one cannot be $100\%$ sure it comes from the $\bar\Sigma(1380)$ resonance, since we do not know the extent to which the $\Sigma(1385)$ is not suppressed. However, the width of $\Sigma(1385)$ (about 40~MeV) is smaller than that of the $\Sigma(1380)$
 (120~MeV as predicted), so the corresponding bump structures should have different widths. On the other hand, the decay into $\pi \bar\Lambda$ should proceed in $s$-wave to give rise to the
quantum numbers of the $\Sigma(1380)$, and proceed in $p$-wave to give rise to the ones of the $\Sigma(1385)$. Thus, the partial wave analysis, if the experimental data is available in future, can also be used to distinguish these two resonances.

In summary, the process $\chi_{c0}\to \bar{\Lambda}\Sigma\pi$ can be used to study the molecular structure of the $\Lambda(1405)$ resonance, and also to test the existences of the predicted states $\Sigma(1380)$ and $\Sigma(1430)$.

\section*{Acknowledgements}
We thank the anonymous referee for the critical comments which were valuable in improving the  presentation of this work. We warmly thank Eulogio Oset for careful reading this paper and useful comments. This work is partly supported by the National Natural
Science Foundation of China under Grant No. 11505158, 11605158, 11475227, 11735003. It is also supported by the Academic Improvement Project of Zhengzhou University.

\end{document}